# Wide range electrical tunability of single photon emission from chromium-based colour centres in diamond

T. Müller[1], I. Aharonovich[2,3], L. Lombez[1], Y. Alaverdyan[1], A. N. Vamivakas[1], S. Castelletto[2,4], F. Jelezko[5], J. Wrachtrup[5], S. Prawer[2] and M. Atatüre[1]

[1]Cavendish Laboratory, University of Cambridge, Cambridge CB3 0HE, UK
[2]School of Physics, University of Melbourne, VIC 3010, Australia
[3]School of Engineering and Applied Science, Harvard University, Cambridge, MA 02138, USA
[4]Centre for Micro-Photonics, Faculty of Engineering and Industrial Science, Swinburne University of Technology, Mail H 34 Hawthorn, VIC 3122, Australia
[5]3. Physikalisches Institut, Universität Stuttgart, 70550 Stuttgart, Germany

**Abstract –** We demonstrate electrical control of the single photon emission spectrum from chromium-based colour centres implanted in monolithic diamond. Under an external electric field the tunability range is typically three orders of magnitude larger than the radiative linewidth and at least one order of magnitude larger than the observed linewidth. The electric and magnetic field dependence of luminescence gives indications on the inherent symmetry and we propose Cr-X or X-Cr-Y type noncentrosymmetric atomic configurations as most probable candidates for these centres.

Room temperature operation of diamond-based colour centres offers a unique platform to the field of quantum photonics[1]. The nitrogen-vacancy (NV) colour centre in diamond has received interest in diverse research fields ranging from spin-based quantum information[2-5] to nanoscale magnetometry[6-9] within the last decade. The impressive level of spin coherence[5], the fast microwave spin manipulation[10], and the recently demonstrated spin-photon interface[11] are essential to the proposed applications. However, the NV centre's broad photon spectrum is unfavourable for many of these experiments. Although emission from the zero-phonon line (ZPL) has been shown to be radiative lifetime broadened, it only accounts for about 4% of the total spectrum, owing to the dominance of phonon-assisted decay. This motivated the search for alternative centres, where potentially a similar degree of spin control could coexist with superior photonic properties. Consequently, silicon-vacancy (Si-V)[12,13] and nickel-based centres, such as NE8[14], were studied in the last few years for this purpose. Recently, chromium-based colour centres were reported among the brightest single-photon emitters in the near infrared spectrum[15,16]. Here, we show that these centres have an impressive level of spectral tunability on the order of a few meVs, while sustaining their oscillator strength, well beyond the range observed for the NV centre. Furthermore, this is the first report of electric field tunability from impurities in diamond other than NV. This electrical tunability is an important asset for generating spectrally indistinguishable single photons from multiple colour centres. In addition, while the NV centre's atomic structure has been known for decades[17], the chromium related defect has been fabricated only recently and even its basic components need to be identified. The particular dependence of the luminescence from chromium-related emitters to an external electric field allows us to infer a non-centrosymmetric atomic configuration.

The colour centres under study were engineered by ion implantation of chromium and oxygen into pure type IIA ([N] < 1 ppm, [B] < 0.05 ppm, Element Six) diamond crystals based on the protocol discussed in Ref. [18]. The implantation energies of the chromium and the oxygen atoms were 50 keV and 19.5 keV, respectively. Based on SRIM simulations such energies will maximize the overlap of these two ions in the diamond lattice. All implantations were performed at room temperature in high vacuum (~$10^{-7}$ Torr). After implantation, the samples were annealed at 1000°C in a forming gas ambient (95% Ar – 5% $H_2$) for 2 hours to repair damage caused by the implantation. The centre is only observed in samples with the right concentration of nitrogen, and co-implanting oxygen (or alternatively sulphur) increased the yield of optically active centres by at least one order of magnitude. All our measurements were carried out at liquid helium temperature (4 K) using a home-built confocal microscope. The defects were optically excited nonresonantly using a continuous wave Titanium:Sapphire laser tuned to 710 nm. The laser light was focused on the sample with a high numerical aperture lens (NA=0.65), which was also used to collect the emission from the defects. Residual pump laser light was filtered out using a 740-nm long pass filter and photoluminescence (PL) was then focused into a single mode fibre which acted as a pinhole. A spectrometer with liquid-nitrogen cooled CCD was used to analyse the PL and time-correlated photon counting was used for lifetime and intensity-correlation measurements. Isolated bright spots on the sample were found by scanning the piezoelectric stage on which the sample was mounted. Electrical gates were fabricated using electron-beam lithography (EBL) and consist of 10 nm of chromium and 40 nm of gold.

Figure 1(a) displays typical photoluminescence of a single chromium-related centre when excited nonresonantly with a 710-nm laser. The spectrum shows spectrometer-resolution-limited (well below 10 GHz) photon emission strongly concentrated in the zero-phonon line (ZPL). A very weak phonon-assisted luminescence can be observed for long timescale integration, which is consistent with the previously reported Debye-Waller factor well exceeding 0.9[16]. Not all centres had resolution-limited ZPL emission and Fig. 1(b) displays an example of broadened emission from a single centre. The excited state lifetime for such centres is around 1 ns[15,18] indicating that the observed broadening is predominantly due to spectral diffusion. Luminescence from all centres studied were linearly polarised to better than 90% as shown in Fig. 1(c). We find that absorption is also linearly polarized and that the emission and the absorption dipoles point along the same direction. These polarisation properties are maintained under the application of both magnetic and electric fields. Figure 1(d) displays typical magnetic field response of these centres up to 7 Tesla external field along the optical axis of the microscope. Within the resolution of our experiments (~5 GHz) the ZPL does not show any shift, splitting, broadening, or loss of emission intensity. These measurements could indicate a transition between states with commensurate Landé factors, as in the case of NV centres. Alternatively, the non-resonant excitation could give rise to spin-polarisation of the centre in a state unaffected by magnetic field. A weak spin-orbit coupling can furthermore make potential transitions to different spin-states undetectable within our signal-to-noise limit, such that no final information can be extracted concerning the spin part of the electronic wavefunction.

The two centres presented in Fig. 1 differ in the actual wavelength of their ZPL emission. Figure 2(a) shows a histogram of single ZPL emission wavelengths from the chromium-based defects grouped in 2-nm bins. The wavelength distribution of observed ZPLs ranges from 740nm up to 800nm, particularly concentrated around 748nm and 775nm. It is interesting to note that while the NV centre exhibits a very well-defined spectral signature, other defects such as the Si-V show a similar inhomogeneous broadening of ZPL centre wavelength[13]. The lack of discrete accumulation of the histogram to a few centre wavelengths suggests that we are not looking at a family of microscopic structures, each having its own ZPL wavelength, but rather the same microscopic structure under a variation of local external effects. For the implanted chromium-related centres, the lattice surrounding each colour centre is quite rich containing a significant number of impurities. The implantation of high doses (~$10^{10}$ ions/cm$^{-2}$) of chromium results in a high chromium concentration in the diamond lattice. In addition, the fabrication process generates a high concentration of substitutional nitrogen, oxygen, vacancies and NV centres while forming a lower density of chromium-based centres[19]. Photoluminescence under a 532-nm laser excitation indeed shows that the NV density is too high to optically resolve individual NV centres. The density of impurity atoms and ions is therefore quite high. We expect the histogram of Fig. 2(a) occurs due to the individual local electric field and strain experienced by each centre due to alien atoms. In addition, the mere presence of a high density of non-carbon atoms can modify the diamond band structure in the vicinity of each centre, contributing to the dispersion we observe in the transition wavelength. One could further speculate that the two suggestive lobes of the histogram may correspond to two charging states of the centre similar to the NV$^0$ and NV$^-$ states, where the charging energy here would be around one third of that for NV centres suggesting slightly weaker confinement. In addition to ZPL wavelength inhomogeneity, most centres show switching between multiple discrete lines and/or dark periods on the order of seconds. Figure 2(b) displays temporal dynamics of the ZPL spectrum for a single centre. The excitation laser power strongly influences these dynamics such that the centres become more stable under lower excitation laser power. This is again consistent with optically induced changes of the charge configuration in the vicinity of each centre, possibly related to photoionisation of the defects in the diamond lattice. Intensity plots along

785.5 nm (red curve) and 786.4 nm (green curve) of Fig. 2(c) display the anticorrelated behaviour of emission originating from the same ZPL. We also note that in rare occasions we observe loss of fluorescence, possibly due to ionisation of the colour centre.

The temporal signature of each centre is unique indicating the high level of sensitivity of these artificially fabricated centres to their local environment. As strain and electric field influence the local surroundings of defects significantly, we proceed to investigate the luminescence of the centres under an externally applied electric field. Emitters are exposed to electric fields using interdigitated gates on the diamond surface, as shown in Fig. 3(a). The width of the fingers, as well as the spacing between two fingers is 3 micrometres. We fabricated structures both parallel to the crystal axes [010] and [100], and at 45$^o$ with respect to the axes. Applying 100 V across the fingers results in local electric field F up to 8.5 x 10$^6$ V/m and the polarity can be reversed by applying a reverse bias. To determine the local field just beneath the diamond surface, we use the Lorentz local electric field approximation F = V($\varepsilon$+2)/2d, with $\varepsilon$ the dielectric constant for diamond, V the voltage applied and d the gate separation. The Stark shift observed in the optical transition should then be given by $h\Delta\nu$= - $\Delta\mu \cdot$ **F** – ½ **F** $\cdot \Delta\alpha \cdot$ **F** in second order perturbation theory, where $h$ and $\Delta\nu$ are Planck's constant and the change in transition frequency. $\Delta\mu$ and $\Delta\alpha$ represent the change in dipole moment and the change in polarisability between the excited and the ground state, respectively.

Figures 3 (b-e) present the response of the ZPL spectrum to the applied electric field for four different centres. A significant part of the observed centres exhibit a linear dependence, as seen in Fig. 3(b) and (c). A predominantly quadratic response was observed for a few centres, as seen in Fig. 3(d) and (e), and about half of the centres remained unchanged by the electric field. The Stark shift (both linear and quadratic) can cause either an increase or a decrease in transition energy when the magnitude of the electric field is increased. Note that the strength and the sign of the observed Stark shift per centre both depend on the respective orientation of the electric dipole and the gate structure. We have investigated 12 emitters with the gate parallel to the [100] axis of the crystal, and 26 emitters with the gate at an angle of 45 degrees with respect to the crystal axis, and we did not find a significant difference in the electric field response between the two directions. For most of the observed centres the linewidths remain largely unaffected under the application of an electric field, as displayed in the inset of Fig. 3(b).

The responses of the chromium-based emitters to electric fields are striking for two reasons. First, the centres exhibit a large tunability of the sharp emission lines, typically two to three orders of magnitude larger than the extracted radiative linewidth and still sustaining the oscillator strength. The range over which the ZPL emission is tunable is at present limited by the maximum applied voltage. This is remarkable and of extreme importance for application of these emitters in generation of indistinguishable photons from different sources[20,21,22]. Using the model presented above, we extract typical linear dipole moments on the order of $\Delta\mu$ = 8.36 GHz/MV/m (1.66 Debye). While this value is comparable to that for NV centres in diamond, the NV centre has a much smaller overall range limited to about ten GHz[23]. With such performance, the investigated chromium-related defect becomes comparable to other solid state systems such as self-assembled quantum dots[24] or conjugated polymer molecules[25] for electrical tunability.

Second, we observe different linear dipole moments for positive or negative field directions. The linear dipole moments can even be effectively zero for one field direction, but finite for the other, as seen in Fig. 3(b). Since we do not observe a change of slope for finite electric field values for either polarity, it is unlikely that this surprising feature is environment mediated, but rather inherent to the centre. As the exact geometry of the present defect is unknown, it is not possible to compare this result with a simple theoretical prediction for a two level system. Nevertheless we can establish an intuitive picture for our findings using an atomic physics model. Considering the level structure of a simple atom in free space, a linear stark shift only occurs for hybridised energy levels due to the well-defined parity of the pure electronic states. We can therefore expect at least one of the probed electronic energy levels of the investigated chromium-related defect to have orbital degeneracy. Furthermore, the heavy dependence of the observed Stark shift coefficient on the electric field polarity for linearly responding centres is only possible for a centre lacking inversion symmetry. This argument can be illustrated by looking at the well-studied NV centre whose symmetry group ($C_{3V}$) does not contain inversion symmetry either. Recent theoretical work on the excited state level structure of the NV centre[26] indicates that an electric field can indeed cause mixing within the excited state manifold leading to eigenstates with different linear dipoles. Assuming a similar structure for the chromium-related centre in the excited (ground) state of the optical transition, and the presence of a relaxation mechanism within the excited state manifold to whichever state has lower (higher) energy for a given electric field, our findings become more plausible. Irrespective of whether this behaviour originates from the ground or the excited state of the optical transition, the lack of inversion symmetry of the defect compound has consequences for the atomic structure of the centre. We can infer that the centre can not consist of a single chromium atom only, since inversion symmetry would not be broken for either a substitutional or interstitial single impurity atom in diamond. Given that the centre is controllably fabricated with a finite yield, we speculate further that the centre is formed by no more than two atomic species and vacancies. We therefore propose a Cr-X or X-Cr-Y type atomic structure for these Cr-based centres, where X and Y are likely to be oxygen, nitrogen or sulphur. We note that one of the atomic species may indeed act purely as a donor or acceptor for a centre nearby. A vacancy for X in Cr-X is unlikely due to the stringent dependence of centre formation on oxygen (sulphur) and nitrogen densities simultaneously. However, either X or Y in X-Cr-Y formation may be occupied by a vacancy still preserving the lack of inversion symmetry. These restrictions on the atomic structure of the centre open the door for simulations of the centre using DFT models to gain further insight into the electronic level structure.

In short, we have observed a large range tunability of chromium-related centres in diamond via the Stark effect. The asymmetric nature of the linear dipole response allows us to infer an atomic configuration lacking inversion symmetry. This most likely leads to an electronic level structure with orbital degeneracy which is lifted under electric fields. Given the strong linear dipole present in the system, charge fluctuations in the environment of the defect will have an impact on the emission properties. Due to the fabrication procedure the chromium centres are embedded in an environment that hosts a number of non-carbon atoms and vacancies. This will lead to not only inherent electric field per centre due to charge distribution, but also a slight bandgap modification for the diamond in the vicinity of the centres. The optical performance of the defects is likely to be significantly enhanced if fabricating defects with a cleaner matrix at the end of the protocol can be achieved, thus making them available to applications in various quantum photonics technologies. Resonant excitation, optically detected magnetic resonance measurements on a single centre, as well as electron spin resonance measurements on an ensemble of centres will shed more light to the level structure under magnetic field.

**Acknowledgments** – We gratefully acknowledge financial support by the University of Cambridge, the European Research Council (FP7/2007-2013)/ERC Grant agreement no. 209636, the Australian Research Council, The International Science Linkages Program of the Australian Department of Innovation, Industry, Science and Research (Project No. CG110039), the European Union Sixth Framework Program under Program No. EQUIND IST-034368, the DFG (SFB/TR21, FOR1482) and the BMBF (EPHQUAM and KEPHOSI). I. A. acknowledges the Grant-In-Aid of Research from Sigma Xi, The Scientific Research Society. We thank C. Matthiesen and C.-Y. Lu for technical assistance, and H. Tureci and D. D. O'Regan for helpful discussions.

**Figure Captions**

**Figure 1. a** Resolution-limited zero phonon line (ZPL) spectrum of a chromium related centre at 4K under nonresonant excitation at 710 nm. The width of the line is well below 10 GHz, and no phonon sidebands can be resolved. **b** An example of a broadened ZPL spectrum of a chromium related centre, with a linewidth of about 60 GHz. Again, phonon sidebands can not be resolved. **c** Magnetic field dependence of the resolution-limited ZPL at 793.1 nm shown in panel a. The magnetic field was applied parallel to the optical axis, and spectra were taken every 500 mT. **d** Polarisation dependence of excitation (blue) and emission (red) of the ZPL transition at 765.1 nm shown in panel b. The solid circles show experimental values and the solid curves are fits to the data with sine square law. The visibility ($I_{max} - I_{min}$)/ ($I_{max} + I_{min}$) is 96.6% for emission and 97.8% for excitation. Emission and excitation are polarised along the same axis.

**Figure 2. a** Histogram of ZPL centre wavelengths of 65 emitters under nonresonant excitation. **b** Temporal behaviour of the ZPL spectrum in 1-s time intervals. Spectral jumps on the order of a nanometre are observed, as well as a temporary loss of fluorescence. **c** Intensity line cuts at 785.5 nm (red) and 786.4 nm (green) of the spectra shown in panel b, displaying the anticorrelated emission intensity of the two lines. The two emission wavelengths do not coexist, but occasionally spectral jumps within the integration time of 1 s lead to reduced contrast.

**Figure 3. a** Interdigitated gold gates on single-crystal diamond. **b** Resolution-limited ZPL emission spectrum as a function of the applied electric field. Negative field values result in a linear dipole coefficient $\Delta\mu$=2.1 GHz/MV/m, positive field values give $\Delta\mu$=0.2 GHz/MV/m. The linewidth of the centre remains unchanged under an electric field, whereas the intensity increases for large negative field values (inset). **c** A centre with a strong linear Stark-shift coefficient $\Delta\mu$=8.36 GHz/MV/m for positive applied field values. The inset shows two spectra taken at 0 MV/m (blue) and 72.8 MV/m (red), with a peak separation of 1.49 nm. At high field values, this centre exhibits line broadening while the overall intensity remains constant. **d** A centre with ZPL at 778.1 nm displaying a predominantly quadratic dependence on the applied electric field with a positive polarisability difference $\Delta\alpha$=0.36 GHz/(MV/m)$^2$ and linear coefficient $\Delta\mu$=0.19 GHz/MV/m. **e** This centre shows a negative polarisability difference $\Delta\alpha$=-0.075

GHz/(MV/m)$^2$, with a linear coefficient $\Delta\mu$=0.054 GHz/MV/m, much weaker than those reported in panel d.

**Figure 1**

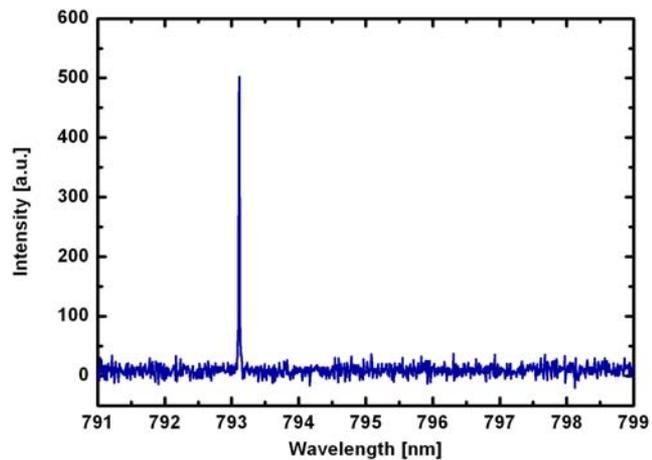
(a)

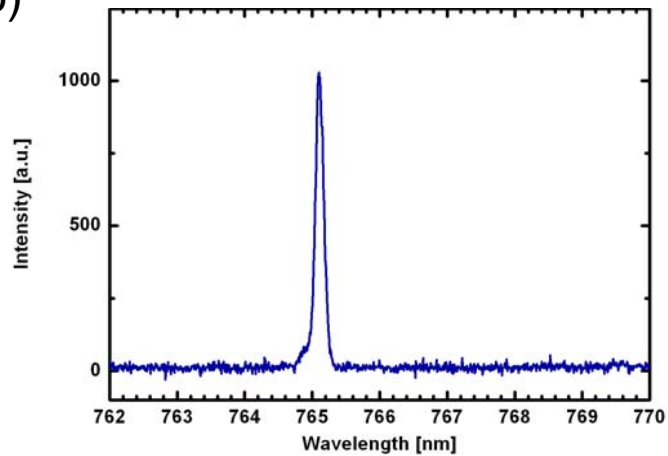
(b)

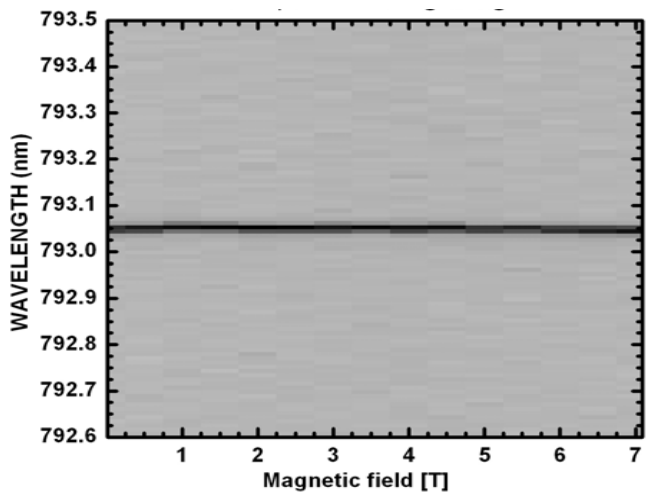
(d)

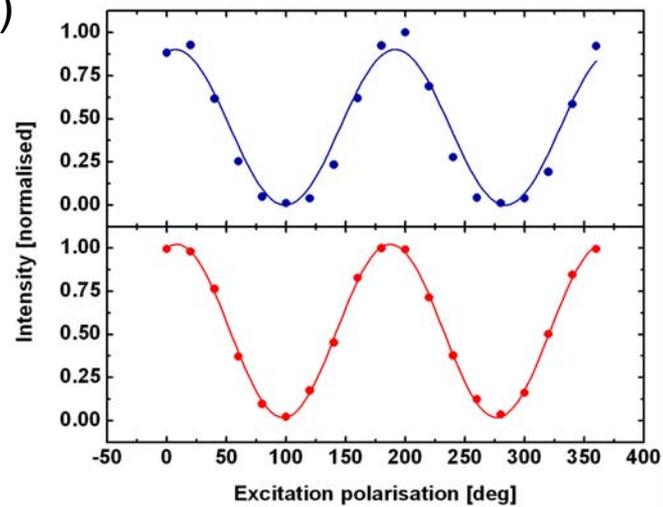
(c)

**Figure 2**

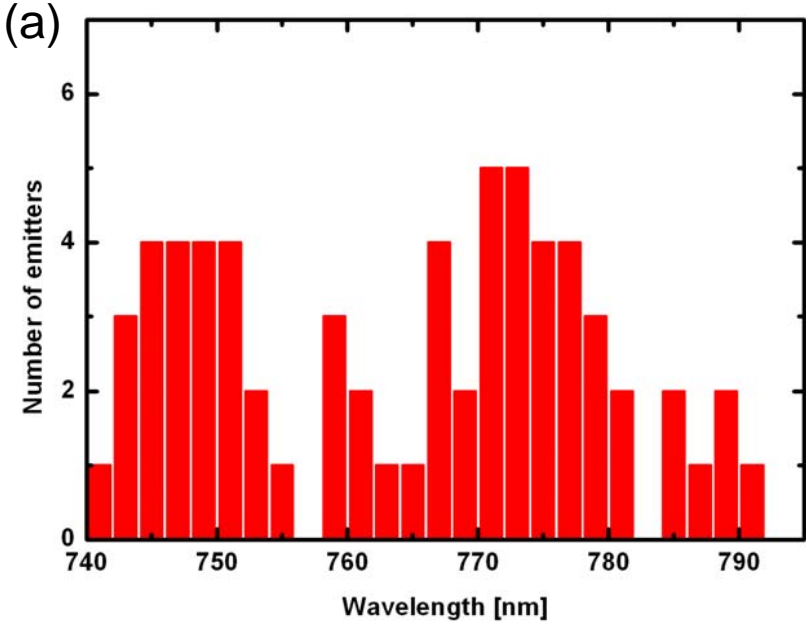
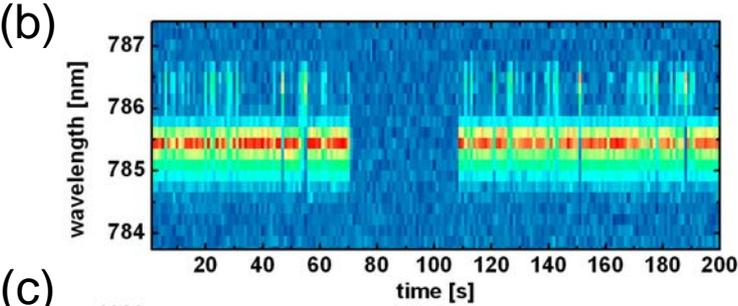
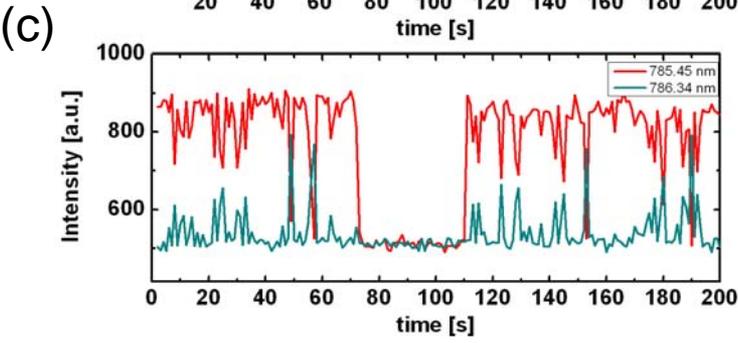

**Figure 3**

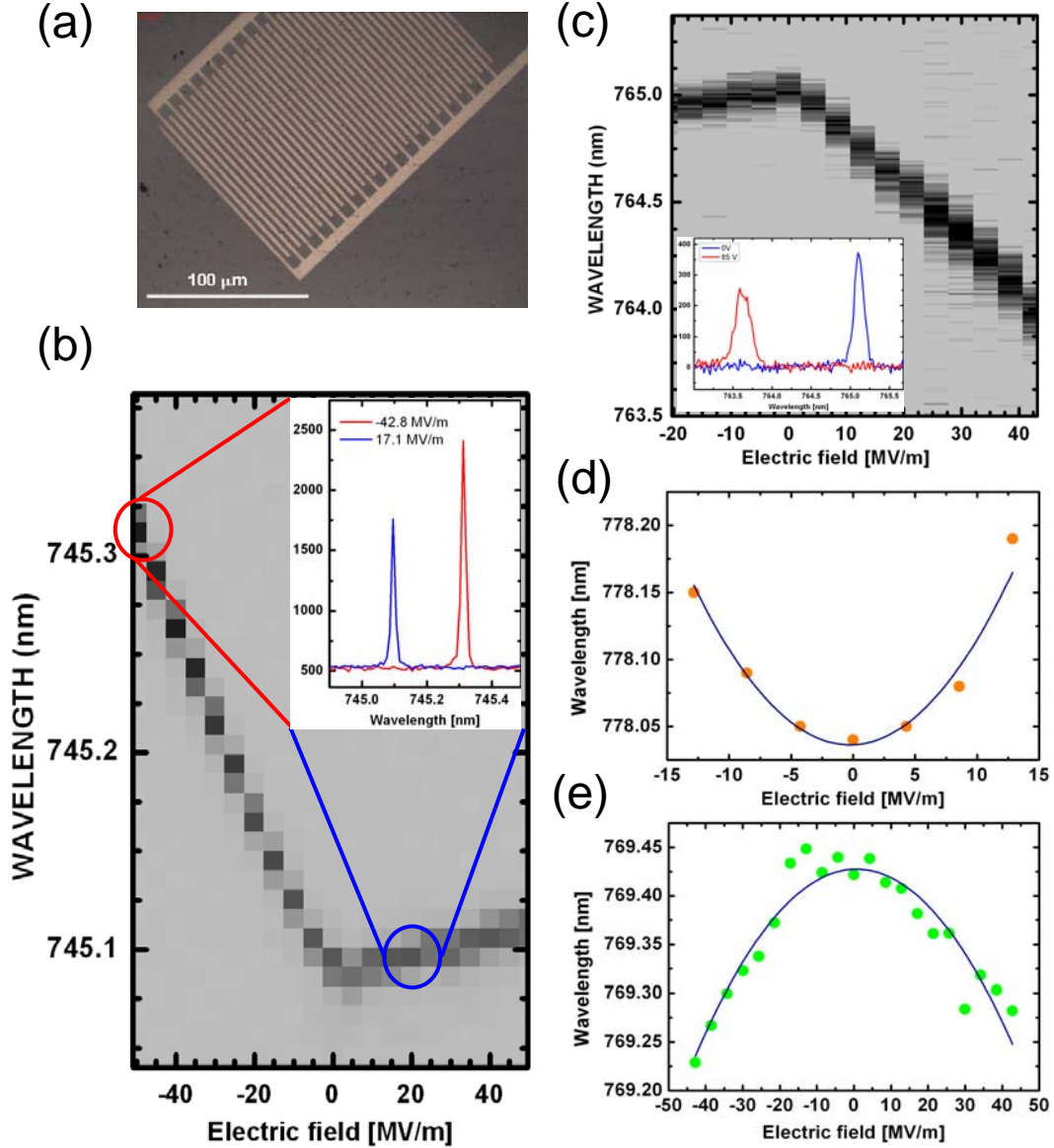